\documentclass[12pt]{article}
\usepackage[utf8]{inputenc}
\usepackage{authblk}
\usepackage[left=2.5cm,right=2.5cm]{geometry}
\usepackage{amsmath}
\usepackage{amssymb}
\usepackage{booktabs}
\usepackage{setspace}
\usepackage{graphicx}
\usepackage{subcaption}
\usepackage{blindtext}
\usepackage[hang,flushmargin]{footmisc}
\usepackage{etoolbox}
\usepackage{pdflscape} 
\usepackage{rotating}  
\usepackage{longtable}
\usepackage{makecell}
\usepackage{multicol}
\usepackage{multirow}
\usepackage{mwe}
\usepackage{gincltex}

\usepackage{lmodern}
\usepackage[dvipsnames]{xcolor}
\usepackage{bigints}
\usepackage[para,online]{threeparttable}
\usepackage{relsize}
\usepackage[final]{pdfpages}
\usepackage{hyperref}

\usepackage[nomarkers, noheads]{endfloat} 

\hypersetup{urlcolor=., colorlinks=false, pdfborder={0 0 0}}  

\providecommand{\keywords}[1]{\textbf{\textit{Keywords:}} #1}



\usepackage{tikz}
\usepackage{tikzscale}
\usetikzlibrary{shapes,decorations,arrows,calc,arrows.meta,fit,positioning, trees, fit}
\tikzset{
    -{Latex[length=2mm, width=2mm]}, ultra thick, 
    auto,node distance =0.8 cm and 0.2 cm, semithick,
    state/.style ={ellipse, draw, minimum width = 0.7 cm},
    point/.style = {circle, draw, inner sep=0.04cm,fill,node contents={}},
    bidirected/.style={Latex-Latex,dashed},
    el/.style = {inner sep=2pt, align=left, sloped},
    mybox/.style={minimum width=4cm,draw,thick,align=center,minimum height=1.8cm}]
}

\title{Making SMART decisions in prophylaxis and treatment studies}

\author[1,2,*]{\small{Robert K. Mahar}}
\author[2,3]{Katherine J. Lee}
\author[4,5,6]{Bibhas Chakraborty}
\author[1,7,8]{Agus Salim}
\author[1]{Julie A. Simpson}
\affil[1]{Centre for Epidemiology and Biostatistics, Melbourne School of Population and Global Health, Faculty of Medicine, Dentistry, and Health Sciences, University of Melbourne, Parkville, Victoria, Australia}
\affil[2]{Clinical Epidemiology and Biostatistics Unit, Murdoch Children's Research Institute, Parkville, Victoria, Australia}
\affil[3]{Department of Paediatrics, Melbourne Medical School, University of Melbourne, Parkville, Victoria, Australia}
\affil[4]{Centre for Quantitative Medicine, Duke-NUS Medical School, Singapore}
\affil[5]{Department of Statistics, Faculty of Science, National University of Singapore, Singapore}
\affil[6]{Department of Biostatistics and Bioinformatics, Duke University, Durham, North Carolina, United States of America}
\affil[7]{School of Mathematics and Statistics, Faculty of Science, University of Melbourne, Parkville, Victoria, Australia}
\affil[8]{Baker Department of Cardiometabolic Health, Melbourne Medical School,Faculty of Medicine, Dentistry and Health Sciences, University of Melbourne, Parkville, Victoria, Australia} 
\affil[*]{Corresponding author: Robert Mahar, Centre for Epidemiology and Biostatistics, Melbourne School of Population and Global Health, 207 Bouverie Street, University of Melbourne, Victoria, 3053
Australia. \texttt{robert.mahar@unimelb.edu.au}}

\begin{document}

\onehalfspacing

\maketitle
\thispagestyle{empty}


\newpage
\onehalfspacing
\begin{abstract}
The optimal prophylaxis, and treatment if the prophylaxis fails, for a disease may be best evaluated using a sequential multiple assignment randomised trial (SMART). A SMART is a multi-stage study that randomises a participant to an initial treatment, observes some response to that treatment and then, depending on their observed response, randomises the same participant to an alternative treatment. Response adaptive randomisation may, in some settings, improve the trial participants' outcomes and expedite trial conclusions, compared to fixed randomisation. But `myopic' response adaptive randomisation strategies, blind to multistage dynamics, may also result in suboptimal treatment assignments. We propose a `dynamic' response adaptive randomisation strategy based on Q-learning, an approximate dynamic programming algorithm. Q-learning uses stage-wise statistical models and backward induction to incorporate late-stage ‘payoffs’ (i.e. clinical outcomes) into early-stage ‘actions’ (i.e.\ treatments). 

Our real-world example consists of a COVID-19 prophylaxis and treatment SMART with qualitatively different binary endpoints at each stage. Standard Q-learning does not work with such data because it cannot be used for sequences of binary endpoints. Sequences of qualitatively distinct endpoints may also require different weightings to ensure that the design guides participants to regimens with the highest utility. We describe how a simple decision-theoretic extension to Q-learning can be used to handle sequential binary endpoints with distinct utilities. Using simulation we show that, under a set of binary utilities, the `dynamic' approach increases expected participant utility compared to the fixed approach, sometimes markedly, for all model parameters, whereas the `myopic' approach can actually decrease utility.  
  
\end{abstract}
\keywords{adaptive trials, adaptive randomisation, Bayesian, COVID-19, decision theory, Q-learning, dynamic programming, sequential multiple assignment randomised trial, Stan, dynamic treatment regimen, adaptive treatment policy}
\thispagestyle{empty}
\newpage
\onehalfspacing


\section{Background} \label{sec:introduction}

Compared to the general population, immunocompromised patients who are exposed to an infectious pathogen may be more likely to become infected, to develop severe disease, and to experience more severe treatment side-effects. Providing these patients first with prophylaxis, to reduce the chance of infection, followed by treatment if subsequently infected, is a common strategy to limit the damage caused by infectious diseases. 

Although the standalone efficacy of prophylaxes and treatments may be well-established, the efficacy of their use \textit{in sequence} may not be. For example, a relatively inefficacious prophylaxis that increases the efficacy of subsequent treatments may be eschewed in favour of a prophylaxis that may be only slightly more efficacious yet decrease subsequent treatment efficacy substantially. Likewise, a later-stage treatment may only be efficacious if a particular prophylaxis precedes it or is given conditional on some particular prognostic prophylaxis outcome being met (for example, disease severity). 

Formalising such prophylaxis and treatment strategies with dynamic treatment regimens is useful. A dynamic treatment regimen is simply an algorithm that specifies sequences of treatment decision rules that depend on time-varying patient characteristics \cite{chakraborty_dynamic_2014, murphy_experimental_2005}. Many dynamic treatment regimens can be candidates for a given clinical condition. The statistical problem is to identify the `optimal' regimen from the candidates under consideration.

Obtaining the data to evaluate the efficacy of a dynamic treatment regimen is best done using a sequential multiple assignment randomised trial (SMART). A SMART is designed to randomise a patient to an initial treatment, observe some response to that treatment and then, depending on the observed response, possibly randomise the same patient to an alternative treatment. Although observational data can be used to optimise dynamic treatment regimens in principle, in practice it is not often done because it is exceedingly difficult to control for time-dependent confounding, among other things \cite{mahar_scoping_2021}.

We are motivated by the study `COVID-19 prevention and treatment in cancer: a sequential multiple assignment randomised trial' (C-SMART) \cite{noauthor_clinicaltrialsgov_nodate}, an ongoing Australian SMART that was designed to evaluate COVID-19 prophylaxis and subsequent treatment in high-risk cancer patients. Although designed as a SMART, C-SMART primarily aimed to estimate standalone prophylaxis and treatment effects, rather than to evaluate dynamic treatment regimens. We consider a stylised version of this trial that has a single initial prophylaxis stage, followed by a single treatment stage, with a primary aim to optimise dynamic treatment decisions. 

In this paper, we extend our study design to include Bayesian response adaptive randomisation at each stage of the SMART. Although non-Bayesian response adaptive designs are possible, Bayesian methods are useful because the interim inferences that are required in an adaptive trial can be made by simply updating the posterior distribution as new data become available. In contrast, frequentist designs often require complex adjustments and are arguably more difficult to interpret.

Because early- and late-stage interventions may interact with one another, a `myopic' response adaptive randomisation strategy, one that ignores between stage information (e.g. outcomes of the early-stage intervention), may lead to an adaptive design that favours suboptimal regimens where only early-stage efficacy is apparent, leading to worse outcomes for trial participants than a fixed design. In principle, this challenge can be overcome by using statistical models for evaluating dynamic treatment regimens as the basis for the response adaptive randomisation, herein referred to as `dynamic' response adaptive randomisation, but few guiding examples exist, and the extent that a `myopic' approach is suboptimal is not known \cite{thall_selecting_2002, thall_covariate-adjusted_2005, cheung_sequential_2015, lee_bayesian_2015, lee_decision-theoretic_2016}. We therefore aim to evaluate a simple two-stage design across a comprehensive range of scenarios to gauge the relative performances of both the `myopic' and `dynamic' response adaptive randomisation approaches, relative to a fixed design.  

We propose to use an approximate dynamic programming method, commonly known as Q-learning \cite{chakraborty_dynamic_2014, watkins_q-learning_1992}, to perform the `dynamic' response adaptive randomisation. Our approach is conceptually similar to that of Lee et al.\, who used dynamic programming to implement decision-theoretic `dynamic' response adaptive randomisation in dose finding studies with joint binary \cite{lee_bayesian_2015} and ordinal \cite{lee_decision-theoretic_2016} intermediate efficacy and toxicity endpoints, and Cheung et al.\ \cite{cheung_sequential_2015}, who used Q-learning to guide response adaptive randomisation in an implementation design with only a single, continuous end-of-study endpoint. 

Two key challenges that arise in evaluating dynamic prophylaxis and treatment regimens are that the endpoints at each stage are both \textit{binary} and also \textit{distinct} from one another, by which we mean that they are qualitatively different at each stage. Intermediate binary endpoints pose a challenge because Q-learning with intermediate outcomes relies on maximising the expectation of a `pseudo-outcome', that is, the sum of the intermediate outcome and the maximised expectation of the later stage outcomes, conditional on patient history. Although the classical implementation of Q-learning is flexible enough to be used when there is either a single end-of-study binary endpoint or sequences of continuous endpoints, sequences of intermediate binary endpoints present a challenge because the pseudo-outcome at the first stage is an awkward sum of a binary endpoint and a probability \cite{moodie_q-learning_2012}. We propose that the approach of Lee et al.\ \cite{lee_bayesian_2015, lee_decision-theoretic_2016} can be adapted to overcome this challenge, thus identifying a methodological connection that has not been made thus far. Further, we propose that the decision-theoretic approach can be used with distinct endpoints, which also pose a challenge because, even if we can deal with intermediate binary endpoints, treating each of these endpoints as equal in utility may be hard to justify in some circumstances. For example, in C-SMART, the endpoints were COVID-19 infection in stage one and all-cause mortality in stage two. This decision-theoretic approach works seamlessly with the proposed Bayesian adaptive design, and thus provides further motivation for the use of Bayesian methods. In contrast to the existing studies, we evaluate the decision-theoretic approach across the entire range of parameter values to better guide practical application and further development of the method. 

This article proceeds by summarising the motivating trial design in Section \ref{sec:design}, describing the Bayesian decision-theoretic design in Section \ref{sec:model} that will allow sequential decisions to be optimised using intermediate and distinct binary endpoints, and describing a simple response adaptive randomisation strategy in Section \ref{sec:oar}. In Section \ref{sec:sims}, we present the results of a simulation study and make concluding remarks in Section \ref{sec:discussion}.

\section{Motivating trial design} \label{sec:design}

C-SMART is a prospective, multi-centre, double-blind, placebo-controlled, Bayesian adaptive SMART. Our stylised version, outlined in Figure \ref{fig:flow}, includes two randomised `substudies' that occur in sequence: a pre-exposure prophylaxis substudy, and a COVID-19 treatment substudy. Participants entering the trial do so within the pre-exposure prophylaxis substudy, where they are randomised to receive either an anti-viral prophylaxis or a placebo with a binary endpoint of a positive COVID-19 diagnosis. If a patient develops COVID-19, then they are randomised to receive either an active COVID-19 treatment or a placebo, with a primary binary endpoint of all-cause mortality. 


\usetikzlibrary{shapes,decorations,arrows,calc,arrows.meta,fit,positioning, trees, fit}
\tikzset{
    -{Latex[length=2mm, width=2mm]}, ultra thick, 
    auto,node distance =0.8 cm and 0.2 cm, semithick,
    state/.style ={ellipse, draw, minimum width = 0.7 cm},
    point/.style = {circle, draw, inner sep=0.04cm,fill,node contents={}},
    bidirected/.style={Latex-Latex,dashed},
    el/.style = {inner sep=2pt, align=left, sloped},
    mybox/.style={minimum width=4cm,draw,thick,align=center,minimum height=1.8cm}]
}

\usetikzlibrary{trees}

\tikzstyle{bag} = [text width=2em, text centered, draw, very thin]
\tikzstyle{end} = [circle, minimum width=3pt,fill, inner sep=0pt]


\tikzstyle{level 1}=[level distance=2cm, sibling distance=1cm]
\tikzstyle{level 2}=[level distance=1cm, sibling distance=5cm]
\tikzstyle{level 3}=[level distance=4.2cm, sibling distance=1.8cm]
\tikzstyle{level 4}=[level distance=1cm, sibling distance=2cm]
\tikzstyle{level 5}=[level distance=4.2cm, sibling distance=1cm]
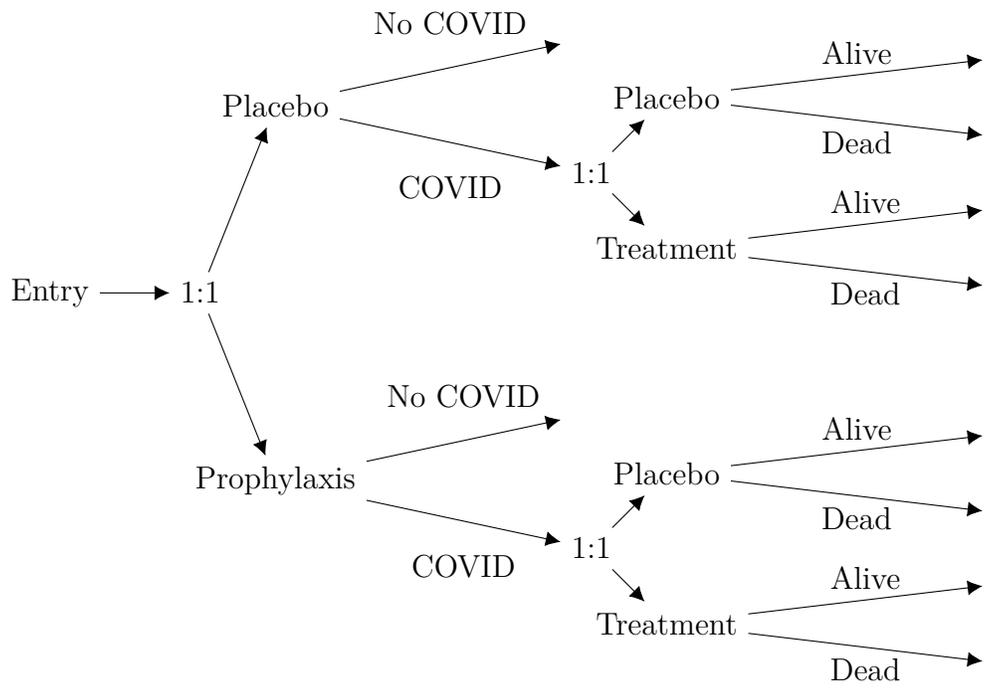
\begin{figure}[htp!]
\caption{C-SMART design} 
\scalebox{1}{
\begin{tikzpicture}[grow=right]
\node {Entry}
    child {
        node {1:1}        
            child {
                node {Prophylaxis}
                child {
                   node {1:1}
                   child {
                     node {Treatment}
                     child {
                       edge from parent
                       node[below]  {Dead}
                     }
                     child {
                       edge from parent
                       node[above]  {Alive}
                     }
                     edge from parent
                   }
                   child {
                     node {Placebo}
                     child {
                       edge from parent
                       node[below]  {Dead}
                     }
                     child {
                       edge from parent
                       node[above]  {Alive}
                     }
                     edge from parent
                   }
                   edge from parent
                   node[below = 0.3 cm]  {COVID}
                }
                child {
                   node[white] {1:1}
                   edge from parent
                   node[above = 0.3 cm]  {No COVID}
                }
                edge from parent
             }
            child {
                node {Placebo}
                child {
                   node {1:1}
                   child {
                     node {Treatment}
                     child {
                       edge from parent
                       node[below]  {Dead}
                     }
                     child {
                       edge from parent
                       node[above]  {Alive}
                     }
                     edge from parent
                   }
                   child {
                     node {Placebo}
                     child {
                       edge from parent
                       node[below]  {Dead}
                     }
                     child {
                     edge from parent
                     node[above]  {Alive}
                     }
                     edge from parent
                   }
                   edge from parent
                   node[below = 0.3cm]  {COVID}
                }
                child {
                node[white] {1:1}
                   edge from parent
                   node[above = 0.3cm]  {No COVID}
                }
                edge from parent
             }
            edge from parent
    };
\end{tikzpicture} 
} 
\centering
\label{fig:flow}
\end{figure}

\section{Decision-theoretic model} \label{sec:model}

\subsection{Sequential actions and their utilities}

Here we describe the general framework for our statistical decision problem \cite{berger_statistical_1980, robert_bayesian_2007}. Let the prophylaxis and disease treatment stages be indexed respectively by $k = 1,2$, from which we define stage-specific actions, $a_k \in \mathcal{A}_k = \{0,1\}$ and binary outcomes, $y_k \in \mathcal{Y}_k = \{0,1\}$.

At each stage we define decision rules $d_k(h_k) \in \mathcal{D}_k$ as functions that map patient history $h_k$ to certain course of action $a_k$ such that $d_k(h_k) \to a_k$, where $h_k$ includes the patient history leading up to, but not including, $a_k$. In our case, we assume that $h_1 = \{\varnothing\}$, that is we do not consider history in the first stage, and $h_2 = \{a_1, y_1\} \times (1 - m)$. Note that  $d_k(h_k)$ will be simply denoted as $d_k$ hereafter. Also note a design constant $m \in \{0, 1\}$ is included to induce either `dynamic' or `myopic' approaches by setting $m = 0$ or $m = 1$ respectively. 

Because the utility of an outcome may depend on what treatments were allocated and patient history, we assign each possible realisation of history, treatment, and outcome at the terminal stage a specific utility, $u(h_k, a_k, y_k)$. The utility of making a decision $d_k$ is defined as the sum of the utilities of each outcome under that decision, weighted by the probability of that outcome. We use a simple utility function that is defined generally as follows:
\begin{equation} \label{eq:general_utility}
    U_k(\pi_k, h_k, d_k, y_k) =  u(h_k, a_k, y_k =0)\times(1-\pi_k) + u(h_k, a_k, y_k = 1)\times\pi_k,
\end{equation}
where $u(h_k, a_k, y_k =0)$ and $u(h_k, a_k, y_k =1)$ are the utilities prescribed to the events where $y_k=0$ and $y_k=1$ for a given pair of $h_k$ and $a_k$, and $\pi_k$ is the probability of an event in stage $k$, conditional on $h_k$ and $a_k$. 

The optimal decisions at each stage, $d^\mathrm{opt}_k$, are defined using dynamic programming, which identifies the optimal stage-specific decision on the basis of the expected utility of that decision and the assumption that all subsequent decisions are optimal. We proceed with a similar approach to Lee et al.\ \cite{lee_decision-theoretic_2016}  by choosing the decision that maximises the expected posterior utility taken with respect to the posterior probabilities of the events at each stage, $p(\pi_k \mid h_k, a_k)$. We use the familiar $Q$-function notation of $Q$-learning to simplify what follows. For any stage, we define a general $Q$-function that represents the posterior expected utility, as follows:
\begin{equation} \label{eq:q_k}
\begin{aligned}
Q_k(\pi_k, h_k, d_k, y_k) & = E_{\pi_k} \Big[U_k(\pi_k, h_k, d_k, y_k) \mid h_k, a_k, y_k \Big]  \\ 
                         & = \int U_k(\pi_k, h_k, d_k, y_k) p(\pi_k \mid h_k, a_k)\ d\pi_k. 
\end{aligned}
\end{equation}
In general, $d_k^\mathrm{opt} = \arg \max_{d_k} Q_k(\pi_k, d_k, h_k, y_k)$ is the optimal stage $k$ decision, conditional on $h_k$.

The stage two utility function is of the same form as \eqref{eq:general_utility}, but with $k=2$,
\begin{equation}
    U_2(\pi_2, h_2, d_2, y_2) =  u(h_2, a_2, y_2 = 0)\times(1-\pi_2) + u(h_2, a_2, y_2 = 1)\times\pi_2.
\end{equation}
The stage one utility function is defined similarly, however the utility of a  stage one infection outcome $y_1 = 1$, is replaced by the \textit{expected} utility under the optimal decision in the second stage, 
\begin{equation} \label{eq:first_stage_utils}
   U_1(\pi_1, h_1, d_1, y_1) =  u(h_1, a_1, y_1 = 0)\times(1-\pi_1) + \max_{d_2} Q_2\Big(\pi_2, h_2, d_2, y_1\Big)\times (1 - m) \times\pi_1.
\end{equation}

In other words, the utility function is the sum of the utility of no infection, weighted by the probability of no infection, and the posterior expected utility of making the likely best treatment decision upon infection, weighted by the probability of developing infection. 

Note also in \eqref{eq:first_stage_utils} the constant $m \in \{0, 1\}$. If $m = 0$ then the second term in the summand in \eqref{eq:first_stage_utils} is unchanged, in other words, all of the later-stage expected utilities are included. If $m = 1$ then the second summand equals zero, in other words, the later-stage expected utilities are not included and the stage-specific utility for the event is zero (i.e. the myopic approach).

\subsection{Probability model}

In general, the posterior distributions defined in \eqref{eq:q_k}, $p(\pi_k \mid h_k, a_k)$ are, by Bayes Theorem, proportional to the product of the likelihood $\mathcal{L}_k(\pi_k, h_k)$ and the prior probability density of $\pi_k$, $p(\pi_k)$. The outcomes $y_k$ are modelled by Bernoulli probabilities with a logistic link and parametric model indexed by stage $k$:
\begin{equation} \label{eq:prob}
\begin{gathered}
    y_k \sim \mathrm{Bernoulli}(\pi_k) \\
    \pi_k = \mathrm{logit}^{-1}\Big[f_k(\beta_k, h_k, a_k)\Big],
\end{gathered}
\end{equation}
where $\pi_k$ denotes the probability of an event $y_k = 1$ and $f_k(\beta_k, h_k)$ is a linear function of history up to stage $k$, and the $\beta_k$ are vectors containing the set of model parameters used for each stage $k$. In general, we can define our posterior distribution as
\begin{equation} \label{eq:posterior}
\begin{aligned}
p(\pi_k \mid h_k, a_k) & \propto \prod_{i\in I_k} \mathcal{L}_k(\pi_k, y_{ik})p(\pi_k) \\
 & \propto \prod_{i\in I_k} f_k(\beta_k, h_k)^{y_{ik}}(1-f_k(\beta_k, h_k))^{1-y_{ik}} p(\pi_k)
\end{aligned}
\end{equation}
where $y_{ik}$ is the specific outcome observed for patient $i \in \{1, ..., N\}$ in stage $k$. 

\section{Response adaptive randomisation} \label{sec:oar}

The overwhelming majority of trial designs randomise between treatments with fixed (and usually equal) probability and only analyse data once the study has been completed. In a SMART design the treatment for patient $i$ at stage $k$ (i.e. $a_{ik}$) is typically allocated randomly with equal probability conditional on $h_{ik}$. This means that, by the end of the trial, many patients will have received sub-optimal treatments. 

Response adaptive randomisation is an approach that aims to improve the probability that trial participants are given the most promising treatment at randomisation by continually reweighting the randomisation ratio in favour of the more promising treatments. In the extreme case of a very poorly performing treatment, this approach can simply stop randomly allocating new patients to that treatment. 

Our approach uses the posterior expected utility to weight the random allocation of treatment in proportion to the total expected utility for the decision being considered:
\begin{equation}  \label{eq:randomise}
     p(a_{ik} \mid h_{ik}) = \frac{Q_k(\pi_k, h_{ik}, d_k = a_{ik}, y_{ik})^c}
     {\mathlarger{\sum}_{\mathcal{D}_k}\Big[Q_k(\pi_k, h_{ik}, d_k, y_{ik})\Big]^c}
\end{equation}
Note the exponent $c$. Where $c = 0$ the probability of random allocation is equal and fixed for all treatments and where $c = 1$ the probability of random allocation is proportional to the relative expected utility of that treatment. 

\section{Simulation study} \label{sec:sims}

Having defined the decision-theoretic model to guide the outcome adaptive randomisation, we proceed by examining the finite-sample properties of our approach using simulated data. The details of the simulation study are provided in the following subsections.

\subsection{Design types}

The four main design types we considered in this simulation study are summarised in Table \ref{tab:settings}. If the design was fixed, then the trial continued until the final outcome was observed at which point the data were analysed. If the design was adaptive, then the trial conducted four scheduled analyses at equal timepoints corresponding to observed outcomes, with randomisation probabilities updated, according to \eqref{eq:randomise}, for each of the first three of the scheduled analyses, before continuing recruitment. If the design was `myopic' then the response adaptive randomisation algorithm ignored between-stage information. If the design was `dynamic' then the response adaptive randomisation algorithm used between-stage information.  

\begin{table}[htp!]  
  \caption{Adaptive and dynamic design settings}
  \hspace{0cm}
  \doublespacing
    \setlength{\extrarowheight}{2pt}
    \begin{tabular}{cr|c|c|c|}
      & \multicolumn{1}{c}{} & \multicolumn{1}{c}{Dynamic}  & \multicolumn{1}{c}{Myopic} \\
      \cline{3-4}     &  Fixed randomisation  & $m = 0, c = 0$ & $m = 1, c = 0$       \\
      \cline{3-4}  & Response adaptive randomisation & $m = 0, c = 1$ & $m = 1, c = 1$  \\
      \cline{3-4}
    \end{tabular}
    \label{tab:settings}
\end{table}

\subsection{Data generation and utilities} \label{sec:data}

Data were randomly generated using the Bernoulli distribution with values of $\pi_k$ as described in Table \ref{tab:utility}. The infection probabilities at the first stage $r_0$, for placebo, and $r_1$, for prophylaxis, were specified from the following set of 21 elements:
\begin{equation}
 R = \{0, 5, 10, ..., 100\}/100,   
\end{equation}
for example, $(r_0,r_1)=(0.1,0.3)$. Likewise, the death probabilities in the second stage either following stage one placebo, $s_0$, or following stage one prophylaxis, $s_1$, were specified from the following set of 8 elements:
\begin{equation}
    S = \{5, 10, 20, 40, 60, 80, 90, 95\}/100,
\end{equation} 
for example, $(s_0,s_1)=(0.45, 0.5)$. Note that there are $21^2=441$ and $8^2=64$ unique pairs of ($r_0,r_1)$ and $(s_0, s_1)$, respectively, resulting in 28224 data generating scenarios.

Also described in Table \ref{tab:utility}, we assumed only a single set of utilities, whereby participants who died received zero utility and participants who lived received unit utility, irrespective of their histories and treatments. Although this is undoubtedly a strong assumption, in life versus death examples it is arguably reasonable to assume that death has no utility and survival is preferred regardless of whether there was infection and subsequent treatment, at least as a demonstrative approximation. This utility set also has the useful property of allowing us to examine the performance of the various design approaches whereby the optimisation problem is reduced to simply maximising the probability of survival.

\begin{table}[htp!] 
\caption{Data generation and utilities}
\centering
\small
\begin{threeparttable}
\begin{tabular}{ 
    >{\centering\arraybackslash}p{0.95cm}  
    >{\centering\arraybackslash}p{1.975cm}
    >{\centering\arraybackslash}p{1.35cm} 
    >{\centering\arraybackslash}p{1.75cm} 
    >{\centering\arraybackslash}p{0.9cm} 
    >{\centering\arraybackslash}p{1.9cm}   
    >{\centering\arraybackslash}p{1.8cm}   
    >{\centering\arraybackslash}p{2.2cm} }
\hline
Stage  & Prophylaxis & Infected & Treatment & Dead  & P(Infected) & P(Dead) & Utility \\ 
$k$   & $a_1$     &  $y_1$     & $a_2$     & $y_2$   &  $\pi_1 \mid a_1$ & $\pi_2 \mid h_2, a_2$  & $u_k(h_k, a_k, y_k)$ \\
\hline
1   & 0      & 0     & --    & --    &  \multirow{5}*{$r_0$}                         & --  & 1 \\
2   & 0      & 1     & 0     & 0     &     & \multirow{2}*{$s_0$}  & 1 \\
2   & 0      & 1     & 0     & 1     &                         &  & 0 \\ 
\cline{7-7} 2   & 0      & 1     & 1     & 0     & &  \multirow{2}*{$s_0$}  & 1 \\ 
2   & 0      & 1     & 1     & 1     &                         &  & 0 \\\hline
1   & 1      & 0     & --    & --    &                    \multirow{5}*{$r_1$}   & -- & 1 \\
2   & 1      & 1     & 0     & 0     &      &  \multirow{2}*{$s_1$}  & 1 \\
2   & 1      & 1     & 0     & 1     &                         &  & 0 \\ 
\cline{7-7} 2   & 1      & 1     & 1     & 0     & &  \multirow{2}*{$s_1$} & 1 \\ 
2   & 1      & 1     & 1     & 1     &                         &  & 0 \\
\hline
\end{tabular}
  \end{threeparttable}
    \begin{tablenotes}
    \footnotesize{Note: $r_0, r_1 \in R: R = \{0, 5, 10, ..., 100\}/100$ and $s_0, s_1 \in S: S = \{5, 10, 20, 40, 60, 80, 90, 95\}/100$.}
    \end{tablenotes}
\label{tab:utility}
\end{table}

We set a maximum sample size of 2,000 patients, which was approximately equal to that of the original C-SMART study, however we note that for our evaluation any sample size could be used. To simplify the simulations, patients were assumed to be recruited at a single time point and their outcomes observed immediately (or within a very short time frame). For each trial design and simulation scenario, we simulated 10 trials. Based on pilot simulations, we judged 10 trials as adequate to allow both the results to be interpreted meaningfully and to complete the required computations within a reasonable amount of time.  

\subsection{Analysis}

For our motivating example, we computed the posterior expected utilities as described in \eqref{eq:general_utility}--\eqref{eq:prob} using the following linear models for \eqref{eq:prob}:
\begin{equation}
\begin{gathered}
f_1(\beta_1, a_1) = \beta_{10} + \beta_{11}a_1 \\ 
f_2(\beta_2, h_2, a_2) = \beta_{20} + \beta_{21}a_2 + \beta_{22}a_1 + \beta_{23}a_1a_2,
\end{gathered}
\end{equation}
where $a_1, a_2 \in \{0,1\}$, and $a_1 = 1$ and $a_1 = 0$, indicates prophylaxis and placebo in the first stage, respectively, and $a_2 = 1$ and $a_2 = 0$, indicates treatment and placebo in the second stage, respectively. 

\subsection{Evaluation}

We evaluated the performance of each four trial designs under each simulation scenario using the empirical mean utility of the trial participants, $\bar{u}$ averaged over the 10 simulated trials, for which we use the shorthand $\bar{\bar{u}}$. We then evaluated the $\bar{\bar{u}}$ of the adaptive-myopic design relative to the fixed-myopic design, and the $\bar{\bar{u}}$ of the adaptive-dynamic design relative to the fixed-dynamic design. 

\subsection{Computational methods} \label{sec:computation}

Markov-chain Monte Carlo (MCMC) methods implemented in \texttt{Stan}, a probabilistic programming language \cite{stan_development_team_stan_2018}, were used to numerically approximate the joint posterior distributions of the parameters for each model. The posterior distribution of each parameter of the analysis models was sampled using the Hamiltonian Monte Carlo algorithm implemented within \texttt{Stan}, called from the \texttt{R} software environment \cite{r_core_team_r_2019}. For each parameter, we ran four MCMC chains in parallel for a ‘warm-up’ phase followed by a ‘sampling’ phase of 1000 iterations each, resulting in 4000 samples from the posterior distribution.

\subsection{Results} \label{sec:results}

\begin{figure}[htp!] 
\caption{Empirical mean utilities of \textit{myopic} SMART with response adaptive randomisation relative to SMART with fixed randomisation}
\centering
  \includegraphics[width=1\textwidth]{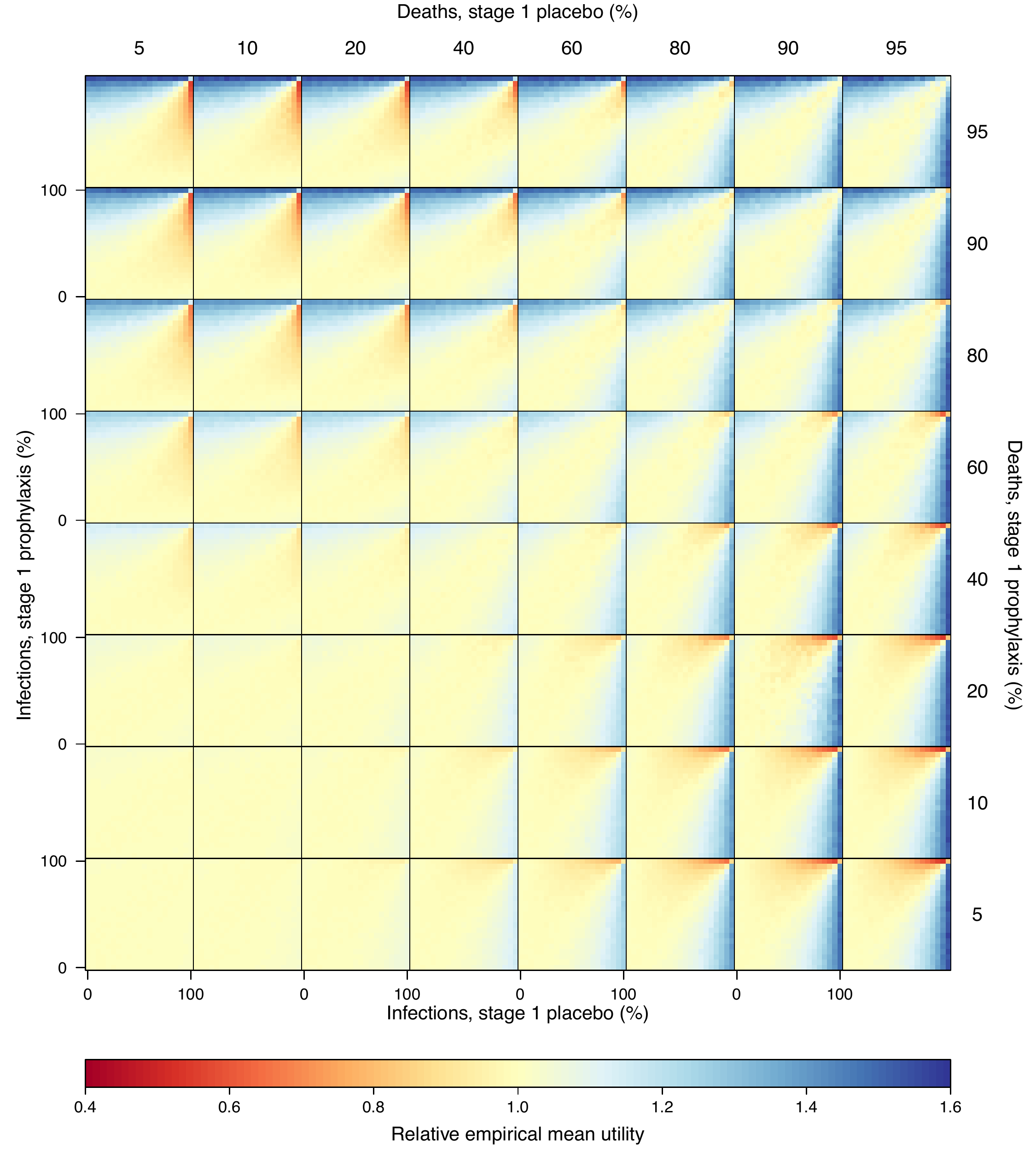}
  \label{fig:rel_utils_myopic}
\end{figure}

\begin{figure}[htp!] 
  \caption{Empirical mean utilities of \textit{dynamic} SMART with response adaptive randomisation relative to SMART with fixed randomisation}
\centering
  \includegraphics[width=1\textwidth]{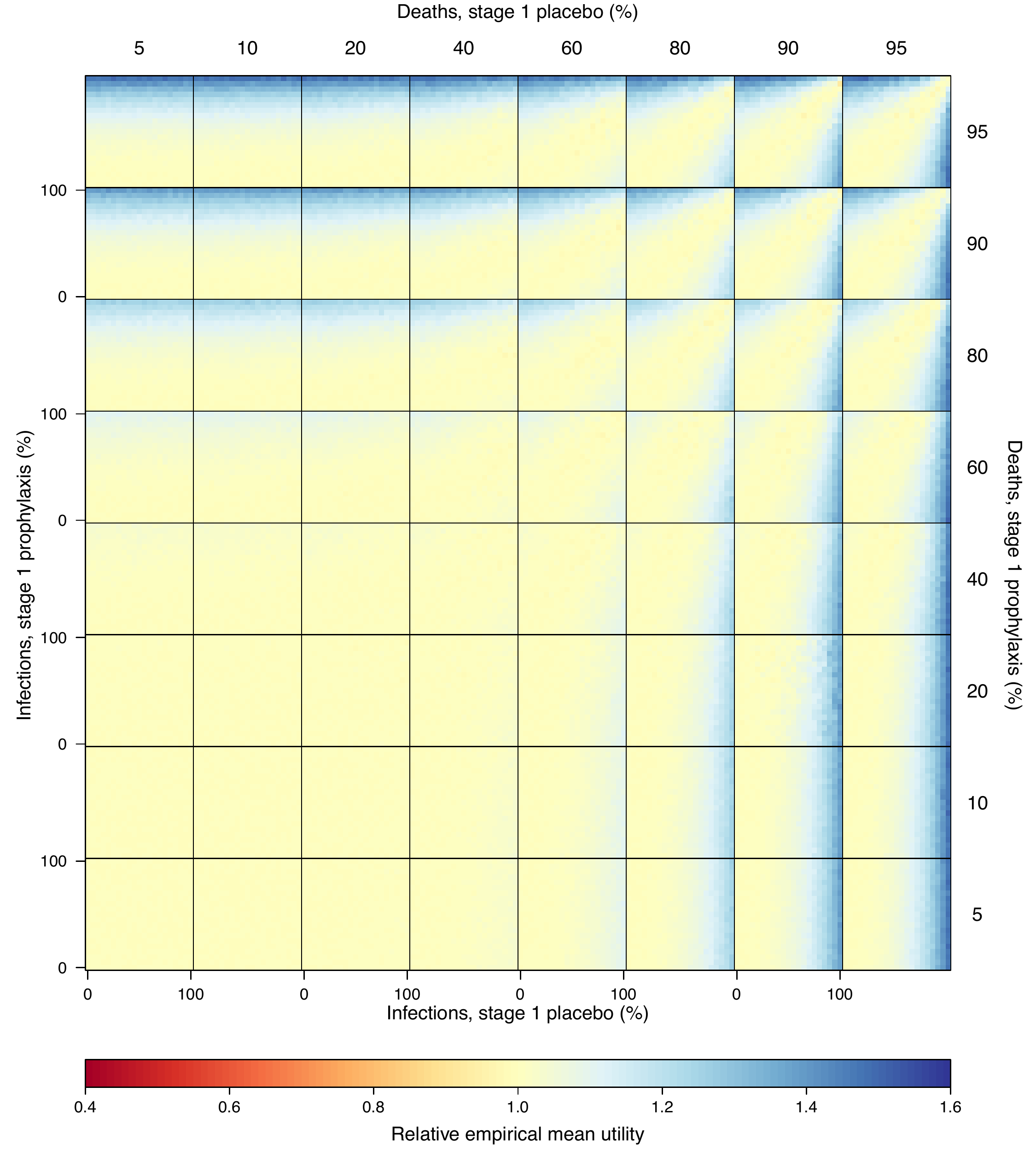}
  \label{fig:rel_utils_dynamic}
\end{figure}

We summarised the results using empirical mean utilities, $\bar{\bar{u}}$, for trial participants (i.e. the average utility of participants, averaged over ten simulations). In each of the figures \ref{fig:rel_utils_myopic} and \ref{fig:rel_utils_dynamic} we show results for all unique 441 combinations of infection probabilities in the placebo arm (inner x-axes) and the prophylaxis arm (inner y-axes), for each of the 64 unique combinations of death probabilities following stage 1 placebo (upper margin) and following stage 1 prophylaxis (right margin). Figure \ref{fig:rel_utils_myopic} shows the mean empirical utilities of the design that used \textit{myopic} response adaptive randomisation, relative to the fixed design. Figure \ref{fig:rel_utils_dynamic} shows the mean empirical utilities of the design that used \textit{dynamic} response adaptive randomisation, relative to the fixed design. Note that where the relative $\bar{\bar{u}}$ was \textit{greater than} one, the response adaptive design had \textit{better} utility than the fixed design, and where the relative $\bar{\bar{u}}$ was \textit{less than} one, the response adaptive design had \textit{worse} utility than the fixed design. Clearly, where the relative $\bar{\bar{u}}$ equalled one, the adaptive and fixed designs had the same utility. 

Myopic response adaptive randomisation favoured the most efficacious prophylaxis, regardless of the subsequent treatment outcomes. In many cases this did not result in relative $\bar{\bar{u}}$ values materially lower than one. For example, if for both prophylaxis histories the infection rates were both very low, or the death rates were similar, then $\bar{\bar{u}}$ was typically close to or greater than one. But where the death rates differed between stage one histories, a very small difference between infection rates could ultimately weight first stage randomisation in favour of a ultimately more fatal treatment regimen and therefore lower utilities, and we see this manifested in areas coloured definitively red in Figure \ref{fig:rel_utils_myopic}. 

Dynamic response adaptive randomisation, by incorporating patient histories, treatments, and expectations about future outcomes in the randomisation algorithm, ensured that the empirical expected utility was either greater than or approximately equal to the fixed design for \textit{all possible data generating scenarios}, as demonstrated by the absence of red coloured areas in Figure \ref{fig:rel_utils_dynamic}. Note the relative utility was approximately equal to one in some scenarios. This is obvious wherever the infection or death probabilities were approximately similar (i.e. $r_0\sim r_1$ and $s_0 \sim s_1$), or where the probabilities and death probabilities were very different (e.g. $r_0 \ll r_1 $ and $s_0 \ll s_1)$. Otherwise, the dynamic response adaptive randomisation algorithm tended to guide participants towards the most efficacious regimens which resulted in, often substantially higher, expected utilities, and this is reflected by the areas coloured blue in Figure \ref{fig:rel_utils_dynamic}.

\section{Discussion} \label{sec:discussion}

We have shown how a Bayesian decision-theoretic dynamic programming algorithm can be used to model intermediate binary endpoints in a simple two-stage prophylaxis and treatment SMART. With a simple binary set of utilities, assigning zero utility to death and unit utility to survival, our approach can be seen as one that simply maximises survival probabilities. Such a utility set may be reasonable in other clinical contexts. Our approach provides a solution to the longstanding challenge for Q-learning of how to model sequential binary endpoints \cite{chakraborty_dynamic_2014, moodie_q-learning_2014}. More complicated and perhaps more relevant SMART trials may now be designed and analysed. Uniquely, our approach allows any utility set to be assigned to different realisations of histories, treatments, and outcomes, allowing decisions based on \textit{distinct} intermediate binary endpoints to be optimised. 
When we use the `dynamic' SMART design with response adaptive randomisation the expected utility of trial participants is either equal or higher, sometimes substantially, to that of a SMART with fixed randomisation. However, we have shown that a `myopic' SMART design with response adaptive randomisation can substantially decrease the expected utility of trial participants relative to a SMART design with fixed randomisation, and for this reason we recommend against using `myopic' response adaptive randomisation for this type of SMART. 

Either the myopic or dynamic approach can easily be extended to multiple stages and/or treatments, and indeed can be applied for SMARTs that re-randomise after all intermediate outcomes, however such designs can become quickly become intractable. Lee et al.\ \cite{lee_decision-theoretic_2016} note that one approach around this intractability is to make a Markovian assumption, in other words, to ignore history. Our `myopic' approach is not strictly Markovian because it also ignores information in later stages, however our results suggest that investigators should proceed with caution if intending to make the Markovian assumption. 


\section*{Data and code availability}

An R package containing all code used for this simulation study is available on online at: \url{https://github.com/robert.mahar/dynamicar}. Alternatively, the R package is available from the authors upon request.


\section*{Acknowledgements}

This work was supported by the Australian National Health and Medical Research Council (NHMRC) Centre for Research Excellence grants to the Victorian Centre for Biostatistics (ID 1035261) and to the Australian Trials Methodology Research Network (ID 1171422). Robert Mahar acknowledges the support of the University of Melbourne Early Career Researcher Grant Scheme and Australian Trials Methodology Research Network Seed Funding Program. Julie Simpson is funded by a NHMRC Investigator Grant (ID 1196068) and  Katherine Lee by a NHMRC Career Development Fellowship (ID 1127984). Bibhas Chakraborty acknowledges start-up funds from Duke-NUS Medical School, Singapore. Research at the Murdoch Children's Research Institute is supported by the Victorian Government's Operational Infrastructure Support Program. We thank the C-SMART investigator team for developing the study design that motivated this work. 

\newpage

\bibliography{references}

\newpage


\end{document}